# Experimental and theoretical evidence for pressure-induced metallization in FeO with the rock-salt type structure

Kenji Ohta,[1,*] R. E. Cohen,[2,*] Kei Hirose,[3,4] Kristjan Haule,[5] Katsuya Shimizu,[1] and Yasuo Ohishi[6]

[1]*Center for Quantum Science and Technology under Extreme Conditions, Osaka University, Toyonaka, Osaka 560-8531, Japan*
[2]*Geophysical Laboratory, Carnegie Institution of Washington, Washington DC 20015, USA*
[3]*Department of Earth and Planetary Sciences, Tokyo Institute of Technology, Meguro, Tokyo 152-8551, Japan*
[4]*Institute for Research on Earth Evolution, Japan Agency for Marine-Earth Science and Technology, Yokosuka, Kanagawa 237-0061, Japan*
[5]*Department of Physics, Rutgers University, Piscataway, New Jersey 08854, USA*
[6]*Japan Synchrotron Radiation Research Institute, Sayo, Hyogo 679-5198, Japan*

Electrical conductivity of FeO was measured up to 141 GPa and 2480 K in a laser-heated diamond-anvil cell. The results show that rock-salt (B1) type structured FeO metallizes at around 70 GPa and 1900 K without any structural phase transition. We computed fully self-consistently the electronic structure and the electrical conductivity of B1 FeO as a function of pressure and temperature, and found that although insulating as expected at ambient condition, B1 FeO metallizes at high temperatures, consistent with experiments. The observed metallization is related to spin crossover.

FeO is one of the fundamental components in the Earth's interior as the iron endmember of ferropericlase, the second most common mineral in the Earth's lower mantle. It is likely to keep the B1 structure throughout Earth's lower mantle according to recent X-ray diffraction studies [1,2]. FeO is insulating under ambient conditions, and is known as a typical Mott or charge-transfer insulator. FeO is a prototypical highly correlated transition metal oxide. It is close to the border between a charge transfer insulator and Mott insulator in the Zaanen-Sawatsky-Allen classification [3], with a large onsite Coulomb repulsion $U$, and smaller charge-transfer energy and oxygen-oxygen hopping, which are both pressure dependent. Along with the other transition metal oxides, it is believed to have a spin crossover or magnetic collapse transition under pressure due to band widening, as predicted by Cohen *et al.* [4]. Investigation of the structure and electrical transport properties of FeO at high pressures and temperatures are of great interest in geophysics as well as condensed-matter physics.

Over two decades ago, the existence of a high-pressure metallic phase of FeO was first suggested based upon measurements of resistivity under shock loading [5,6], and the observed metallization has long been considered to be due to a structural transition to the NiAs (B8) structure [7]. In this Letter, we provide evidence for a metal-insulator transition in FeO at high temperature and pressure within the B1 structure from *in-situ* high *P-T* electrical resistance measurements and fully self-consistent electronic structure computations using Density Functional Theory-Dynamical Mean Field Theory (DFT-DMFT) with continuous time quantum Monte Carlo (CTQMC).

We performed simultaneous electrical resistance and X-ray diffraction (XRD) measurements on FeO *in situ* at high *P-T* conditions in a laser-heated diamond-anvil cell (DAC) with a membrane system. The sample was fine powdered $Fe_{0.96}O$ and we used beveled 120-μm culet diamond anvils. The disk of sample and the gold electrodes were sandwiched between $SiO_2$ glass layers in a sample chamber at the center of electrically insulating gasket that consists of rhenium and cBN powder. The sample was heated in a double-sided heating system with a fiber laser. The electrical resistance of sample was measured at high *P-T* conditions using a quasi-four-terminal method, concurrently with XRD measurements to determine the crystal structure of FeO. Measured resistance using the quasi-four-terminal method includes the resistance of gold electrodes, but the contribution of gold resistance is small relative to the resistance of FeO (see supplemental material [8]). This procedure is the same as that employed in our previous study [9]. Pressurization in a DAC was conducted by gas charging into the membrane system, which enables us to compress the sample during laser heating. Pressures were determined from the unit-cell volume of gold (electrode) obtained by the XRD measurements, using its *P-V-T* equation of state [10]. The electrical conductivity of B1 FeO was estimated from the resistance of FeO and the sample geometry that is defined by the distance between the electrodes, the size of the laser spot, and the thickness of the sample [11]. Each run was carried out after thermal annealing that reduced the deviatoric stress in the sample.

We conducted three separate runs in a pressure range from 32 to 141 GPa (Fig. 1). It is known that B1 FeO undergoes a second order phase transformation into the rhombohedrally distorted B1 (rB1) phase below the Néel temperature [12,13], and shows a further phase change to the B8 structure at higher pressure [1,7,14,15].

Recently, a CsCl-type (B2) phase of FeO was found to be exist above 240 GPa and 4000 K [16].

The first experiment was carried out between 32 and 132 GPa at high temperatures (circle symbols in Fig. 1). Between 30 and 50 GPa, XRD spectra show the structure change from rB1 to B1 with increasing temperature. The resistance of the rB1 phase dramatically decreased with increasing temperature, as is expected in an insulator. The resistance of B1 FeO showed a much smaller temperature dependence [Fig. 2(a), (b)], consistent with being a bad metal or bad insulator, i.e. intermediate between prototypical metallic and prototypical insulating behavior. The observed non-metallic behavior in rB1 and B1 FeO is in good accordance with that obtained in our previous study [9]. We next measured the resistance from 58 GPa and 300 K to 73 GPa and 2270 K after gas compression [Fig. 2(c)]. The temperature dependence of the B1 resistance changed sign to positive at 70 GPa and 1870 K. The positive temperature slope is consistent with metallic behavior; we find that B1 FeO metallizes at that $P$-$T$ condition. We further measured the resistance of B1 FeO at higher pressures up to 132 GPa and 2320 K, indicating it remained metallic [Fig. 2(d)]. We obtained a temperature coefficient ($\alpha$; $\rho(T) = \rho(T_0)\{1+\alpha(T-T_0)\}$, where $\rho$, $T$ and $T_0$ are electrical resistivity, temperature and reference temperature, respectively) of metallic B1 FeO of $(3.2 \pm 0.3) \times 10^{-4}$ K$^{-1}$, which did not change appreciably with pressure [Fig. 2(d)]. In the second and third sets of experiments, we also observed metallization of B1 FeO, confirming the first set of experiments (Fig. 1). The present results demonstrate that the metal-insulator transition in B1 FeO occurs at around 70 GPa and 1900 K. The transition boundary has a negative *P-T* slope, which was determined from our data in a temperature range between 1400 and 2000 K (Fig. 1). Throughout all the experimental runs, no evidence for reaction or decomposition of FeO was not observed from obtained XRD spectra.

Knittle *et al.* [5] first reported the metallization of $Fe_{0.94}O$ under shock-wave compression. They observed high electrical conductivity of FeO approximately of $10^6$ S/m comparable to that of pure iron and iron-silicon alloy above 72 GPa. They observed a decrease in the conductivity with increasing shock compression, and thus higher temperatures, which also was evidence for metallization. It was thought that this metallization corresponds to the transition to the B8 structure [7] but it now appears that the B8 structure does not appear until higher pressures at these temperatures, and the metallization we observe occurs in the B1 structure at high temperatures. Electrical conductivity of metallic B1 phase measured in this study is much lower than $10^6$ S/m, although positive temperature dependence of the B1 resistance obviously indicates the metallic nature. The discrepancy in the resistivity between present and previous measurements could be derived from variant chemical compositions in FeO ($Fe_{0.94}O$; Knittle *et al.* [5], $Fe_{0.96}O$; this study). Indeed, the electrical conductivity of $Fe_{0.91}O$ is twice as high as that of $Fe_{0.94}O$ at 1 bar and low temperatures [17].

Our theoretical calculations also show metallization, are consistent with our experimental observations, and reveal the mechanism of metallization of B1 FeO. In the DFT-DMFT method [18], the strong correlations on Fe ion are treated by the DMFT, adding self-energy $\Sigma(i\omega)$ to the DFT Kohn-Sham Hamiltonian. The self-energy $\Sigma(i\omega)$ contains all Feynman diagrams local to the Fe ion. No downfolding or other approximations were used, and the calculations are all-electron as implemented in Ref [19]. The self-consistency matrix equation is $P(i\omega + \mu - H^{KS} - E\Sigma')^{-1} = (i\omega - E_{imp} - \Sigma - \Delta)^{-1}$, where $P$ is the projection from the crystal with the LAPW representation to the Fe local orbitals, $m$ is the chemical potential adjusted to get the right number of electrons, $H^{KS}$ is the Kohn-Sham DFT Hamiltonian, $E$ is the embedding of the impurity into the crystal (inverse of $P$), $\Sigma'=\Sigma-E_{DC}$ where $E_{DC}$ is the double counting correction, and $E_{imp}$ and $\Delta(i\omega)$ are the impurity levels and hybridization, respectively. The impurity solver takes as input $E_{imp}$ and $\Delta(i\omega)m$ and delivers $\Sigma(i\omega)$ as the output. We used the Wu-Cohen GGA exchange correlation functional in $H^{KS}$ [20]. Brillouin zone integrations were done over 1000 k-points in the whole zone in the self-consistent calculations and 8000 k-points for the density of states and conductance computations. The impurity model was solved using Continuous Time Quantum Monte Carlo (CTQMC) [21,22]. On the order of 100 DFT and DMFT cycles were required for self-consistency. Calculations were fully self-consistent in charge density, chemical potential and impurity levels, the lattice and impurity Green's functions, hybridizations and self-energies. The densities of states and conductivities were computed from analytic continuation of the self-energy from the imaginary frequency axis to real frequencies using an auxiliary Green's function and the maximum entropy method, taking care that the zero frequency limit of imaginary and real axis self-energies agree. The reported conductivities are the low energy limit of the optical conductivity.

We computed the electronic structure of cubic B1 FeO at 300 and 2000 K, and high pressures. At room temperature, we obtain an insulating state with a gap [Fig. 3(a), (b)]. At high temperatures of 2000 K a pseudogap forms at 13 GPa, and we find FeO to be a bad metal with a very low conductivity [Fig. 3(c)]. With compression the gap closes and the density of state is metallic [Fig. 3(d)]. We see a broad high-spin to low-spin crossover starting at 70 GPa and finishing about 200 GPa in the local spin susceptibility and in the $e_g$ and $t_{2g}$ occupancies.

At 300K, we find the following: a cubic paramagnetic insulator in a local high spin state at low pressures. At about 70 GPa and 300 K we find a crossover to a low spin state, becoming a low-spin insulator. There is a small pressure range that is high-spin metallic at about 70 GPa, but it becomes a low-spin insulator at higher pressures, finally metallizing at a compression of a factor of two, and a pressure of about 220 GPa. We find an enhanced thermal expansivity $\alpha$ in the metallic phase, consistent with the thermodynamic relationship $\alpha = \dfrac{\gamma C_V}{K_T V}$ with the addition of electronic heat capacity $C_V^e$ to that of phonons, and the electronic contribution to the Grüneisen parameter $\gamma^e = \dfrac{d \ln g(E_F)}{d \ln \rho}$, the log derivative of the electronic density of states (DOS) at the Fermi level with density; our calculations show that the DOS increases with compression up to 200 GPa so that the electronic contribution to $g^e$ is positive. The high temperature computed conductivity is compared with experiment in Fig. 4. We find excellent agreement, especially considering the difficulty of estimating the exact sample geometry in the experiments, and neglect of phonons and defects in the computations. We find that the high conductivity at 70 GPa and above is due to the underlying spin transition; the metallicity at high temperatures is due to thermal fluctuations between the high and low spin states enhanced by the presence of a wide *4s* band just above the Fermi level. At low temperatures the small metallic region between high-spin and low-spin also has large quantum fluctuations between high and low spin configurations, again leading to metallic behavior. Interestingly the metallization at low temperatures is consistent with that found by Gramsch *et al.* [23] for LDA+U with the preferred U of 4.6 eV for the strained ground state monoclinic structure.

Our calculations do not agree with the DMFT computations of Shorikov *et al.* [24] who found metallization at low temperatures in FeO at 60 GPa persisting to over 140 GPa with no spin crossover. Their computations were restricted to Fe *3d* orbitals only (downfolded), and the calculations were not charge self-consistent. These approximations are likely the reason for the difference in the results. Shorikov *et al.* [24] claim agreement with the metallization observed by Knittle *et al.* [5] but neglected the fact that latter experiments were performed at high temperatures.

Struzhkin *et al.* [25] also observed possible metallization in FeO at ambient temperatures at megabar pressures. It is not known whether their sample converted to the B8 structure stable under those conditions or not, but it could have been the B1 (or rB1) phase since at room temperature rB1 is general preserved metastably in the stability field of B8 phase [26,27]. The high-spin metallic region we find may be consistent with those experiments, and lattice strain, magnetic ordering, and non-stoichiometry could shift or broaden the range of metallization. More recently Ozawa *et al.* [27] investigated the relation between crystal structure and spin state of FeO at room temperature after laser heating. They showed that high-spin rB1 FeO transformed at 100 GPa into inverse B8 phase with high spin state that may be insulator, and then underwent normal B8 phase with low-spin state and metallic nature at around 120 GPa. Just recently Fischer *et al.* [28] presented measurements of emissivity of FeO at high pressures and temperatures that show metallization consistent with our results

FeO adopts the metallic B1-type phase in the Earth's lowermost mantle and the top of outer core conditions (Fig. 1), and it could exist there [29-34]. Electrical conductivity of metallic B1 FeO obtained in this study is about $9.0 \times 10^4$ S/m at 135 GPa and 3700 K, corresponding to the conditions at the core-mantle boundary [e.g., 35], which is much higher than those of natural mantle materials such as pyrolitic mantle [11,36]. Presence of such highly conductive FeO at the core-mantle boundary region can enhance the electromagnetic interaction between solid mantle and liquid core, which would induce the anomalous features in observed Earth's rotation [32,37]. Finally, since we know that the MgO endmember of magnesiowüstite is insulating throughout the Earth, the existence of metallic FeO requires a two-phase field for the MgO-FeO binary system. This will modify the MgO-FeO-$SiO_2$ ternary for iron rich compositions, so that phase relations in the deep Earth could be more complicated than assumed.

We thank S. Tateno, H. Ozawa, and H. Gomi for support in the experiments and thank G. Kotliar for helpful discussions. The synchrotron XRD measurements were conducted at the BL10XU of SPring-8 (proposal no. 2011A0099). Computations were performed on the NSF teragrid machine longhorn and at the Carnegie Institution. K.O. is supported by the Japan Society for Promotion of Science Research Fellowship for Young Scientists. R.E.C. was supported by the Carnegie Institution.

*To whom correspondence should be addressed: ohta@djebel.mp.es.osaka-u.ac.jp (K.O.) and rcohen@ciw.edu (R.E.C.)

[1]  H. Ozawa, K. Hirose, S. Tateno, N. Sata, and Y. Ohishi, Phys. Earth Planet. Inter. **179**, 157 (2010).

[2]  R. A. Fischer *et al.*, Earth Planet. Sci. Lett. **304**, 496 (2011).

[3]  J. Zaanen, G. A. Sawatzky, and J. W. Allen, Phys. Rev. Lett. **55**, 418 (1985)

[4]  R. E. Cohen, I. I. Mazin, and D. G. Isaak, Science **275**, 654 (1997).

[5]  E. Knittle, R. Jeanloz, A. C. Mitchell, and W. J. Nellis, Solid State Comm. **59**, 513 (1986).

[6]  E. Knittle, and R. Jeanloz, Geophys. Res. Lett. **13**, 1541 (1986).

[7]  Y. Fei, and H. K. Mao, Science **266**, 1678 (1994).

[8]  See supplemental material for detail.

[9]  K. Ohta, K. Hirose, K. Shimizu, and Y. Ohishi, Phys. Rev. B. **82**, 174120 (2010).

[10]  K. Hirose, N. Sata, T. Komabayashi, and Y. Ohishi, Phys. Earth Planet Inter. **167**, 149 (2008).

[11]  K. Ohta *et al.*, Science **320**, 89 (2008).

[12]  G. Zou, H. K. Mao, P. M. Bell, and D. Virgo, Carnegie Inst. Washington Yearb. **79**,

374 (1980).

[13]   T. Yagi, T. Suzuki, and S. Akimoto, J. Geophys. Res. **90**, 8784 (1985).

[14]   I. I. Mazin, Y. Fei, R. Downs, and R. Cohen, Am. Mineral. **83**, 451 (1998).

[15]   T. Kondo, E. Ohtani, N. Hirao, T. Yagi, and T. Kikegawa, Phys. Earth Planet. Inter. **143-144**, 201 (2004).

[16]   H. Ozawa, F. Takahashi, K. Hirose, Y. Ohishi, and N. Hirao, Science **334**, 792 (2011).

[17]   D. S. Tannhauser, J. Phys. Chem. Solids **23**, 25 (1962).

[18]   G. Kotliar *et al.*, Rev. Mod. Phys. **78**, 865 (2006).

[19]   K. Haule, C. H. Yee, and K. Kim, Phys. Rev. B **81**, 195107 (2010).

[20]   Z. Wu, and R. E. Cohen, Phys. Rev. B **73**, 235116 (2006).

[21]   K. Haule, Phys. Rev. B. **75**, 155113 (2007).

[22]   E. Gull *et al.*, Rev. Mod. Phys. **83**, 349 (2011).

[23]   S. A. Gramsch, R. E. Cohen, and S. Y. Savrasov, Am. Mineral. **88**, 257 (2003).

[24]   A. O. Shorikov, Z. V. Pchelkina, V. I. Anisimov, S. L. Skornyakov, and M. A. Korotin, Phys. Rev. B **82**, 195101 (2010).

[25]   V. V. Struzhkin *et al.*, Mater. Res. Soc. Symp. Proc. **987**, 0987-PP05-04 (2007).

[26]   J. Badro *et al.*, Phys. Rev. Lett. **83,** 4101 (1999).

[27]   H. Ozawa *et al.*, Phys. Rev. B **84**, 134417 (2011).

[28]   R. A. Fischer *et al.*, Geophys. Res. Lett.*,* doi:10.1029/2011GL049800, in press.

[29]   D. P. Dobson, and J. P. Brodholt, Nature (London) **434**, 371 (2005).

[30]   S. Labrosse, J. W. Hernlund, and N. A. Coltice, Nature (London) **450**, 866 (2007).

[31]  R. Nomura *et al*., Nature (London) **473**, 199 (2011).

[32]  B. A. Buffett, E. J. Garnero, and R. Jeanloz, Science **290**, 1338 (2000).

[33]  B. A. Buffett, and C. T. Seagle, J. Geophys. Res. **115**, B04407, (2010).

[34]  G. Helffrich, and S. Kaneshima, Nature (London) **468**, 807 (2010).

[35]  S. Tateno, K. Hirose, N. Sata, and Y. Ohishi, Earth Planet. Sci. Lett. **277**, 130 (2009).

[36]  K. Ohta *et al*., Earth Planet. Sci. Lett. **289**, 497 (2010).

[37]  R. Holme, in *The Core-Mantle Boundary Region, 1998,* edited by M. Gurnis, M. E. Wysession, E. Knittle, and B. A. Buffett (American Geophysical Union, 1998), p. 139.

[38]  R. Boehler, Earth Planet. Sci. Lett. **111**, 217 (1992).

**Figure captions**

FIG. 1. Phase diagram of FeO. Stabilities of rB1, insulating B1, and metallic B1 phases are represented by solid, gray solid and open symbols, respectively. Circles, squares and triangles indicate each set of experiments (Run#1~3). A metal-insulator transition boundary shown as bold line is determined from present data, and linearly extrapolated to the melting condition (broken bold line). The estimated uncertainty in location of the transition is shown by gray band. The melting curve and the phase boundaries of FeO shown as broken lines are from previous studies [1,7,38]. The uncertainty in temperature was about ±10 %, and that in pressure was smaller than ±5 GPa, mainly due to the variation in temperature when the equation of state of gold was applied.

FIG. 2. Variations in measured resistance of rB1 (solid), insulating B1 (gray solid) and metallic B1 (open) phases as a function of temperature in the first run. Those were measured in a *P-T* range (a) from 32 GPa and 300 K to 41 GPa and 1870 K, (b) from 44 GPa and 505 K to 53 GPa and 1960 K, (c) from 58 GPa and 300 K to 73 GPa and 2270 K, and (d) from 76 GPa and 1330 K to 132 GPa and 2230 K. The pressure values shown in (d) are those calibrated at about 1800 K. Accuracy of the resistance is within ±0.01 %.

FIG. 3 (color). Densities of states (DOS) at 300 K and 2000 K at two volumes, 540 bohr$^3$ and 405 bohr$^3$. The density of states was computed from the DFT-DMFT results. Pressure values were determined from *P-V-T* equation of state of B1 FeO [2]. (a) There is a gap at ambient conditions (the small DOS in the gap is numerical from the analytic continuation). The gap is of Mott and charge transfer character, having both Fe *d* and O *p* states on both sides on the gap. (b) Under pressure (68 GPa) a high-spin to low-spin transition occurs, as can be seen from the decrease in $e_g$ and increase in $t_{2g}$ occupancies (DOS below the Fermi level $E_F$ at 0). (c) At high temperatures at low compression (13 GPa and 2000 K) the gap turns into a pseudogap, and FeO is a bad metal. (d) At high temperatures and higher pressures (88 GPa and 2000 K) FeO is a good metal with no gap, or even "filled gap" at $E_F$.

FIG. 4. Electrical conductivities of B1 FeO determined by experiment and theory as a function of pressure. Open squares, experimental data measured at about 1850 K; solid circles, theoretical results calculated at 2000 K. The errors in conductivity measurements were derived mainly from the uncertainty in the sample thickness, which should be smaller than ±25%.

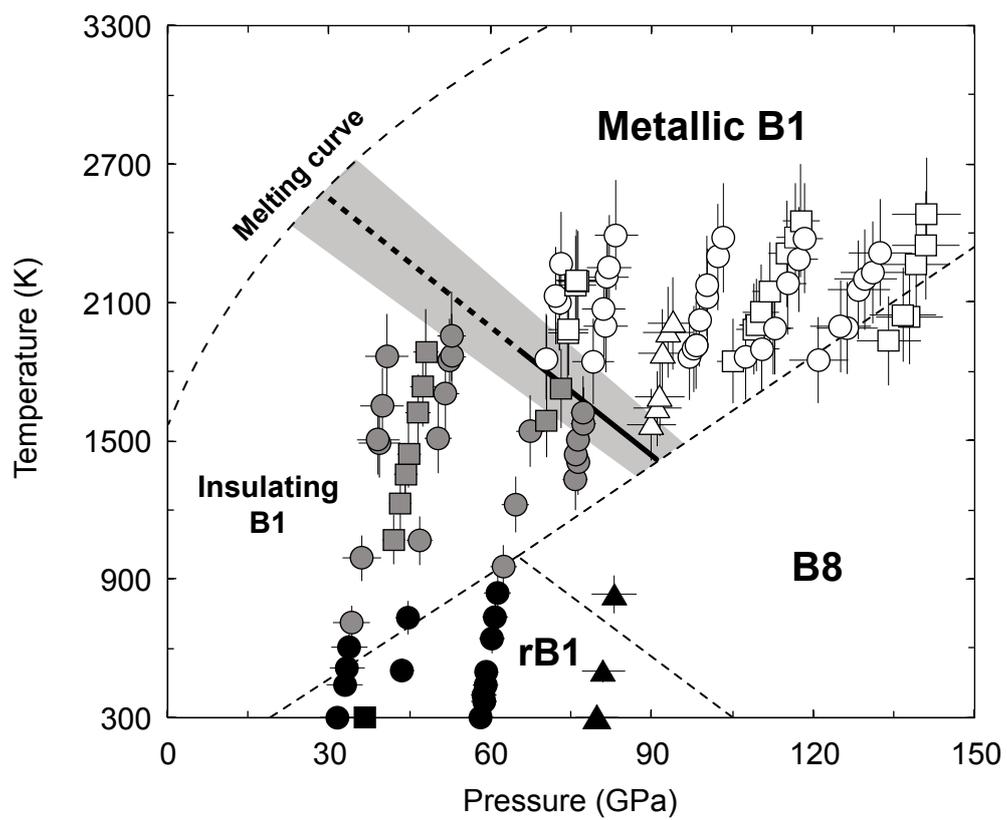

FIG. 1

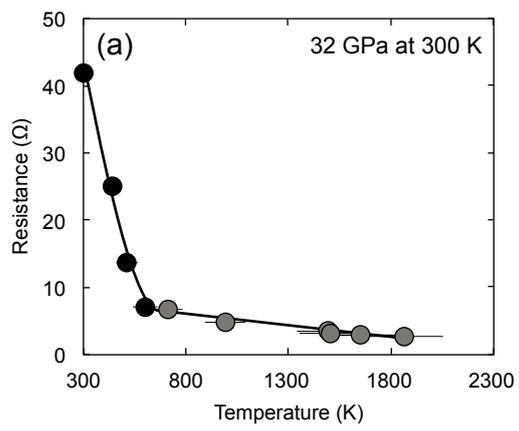
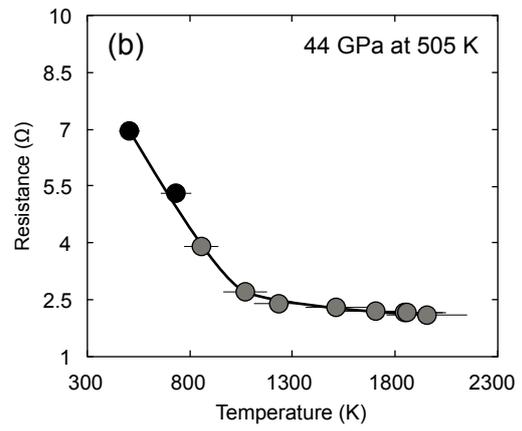
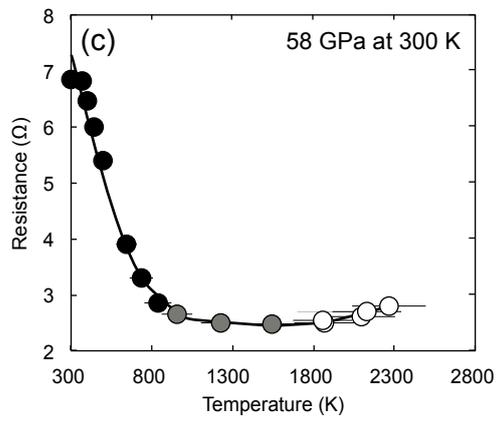
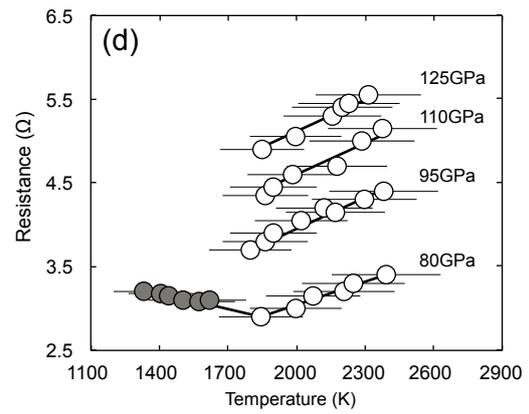

FIG. 2

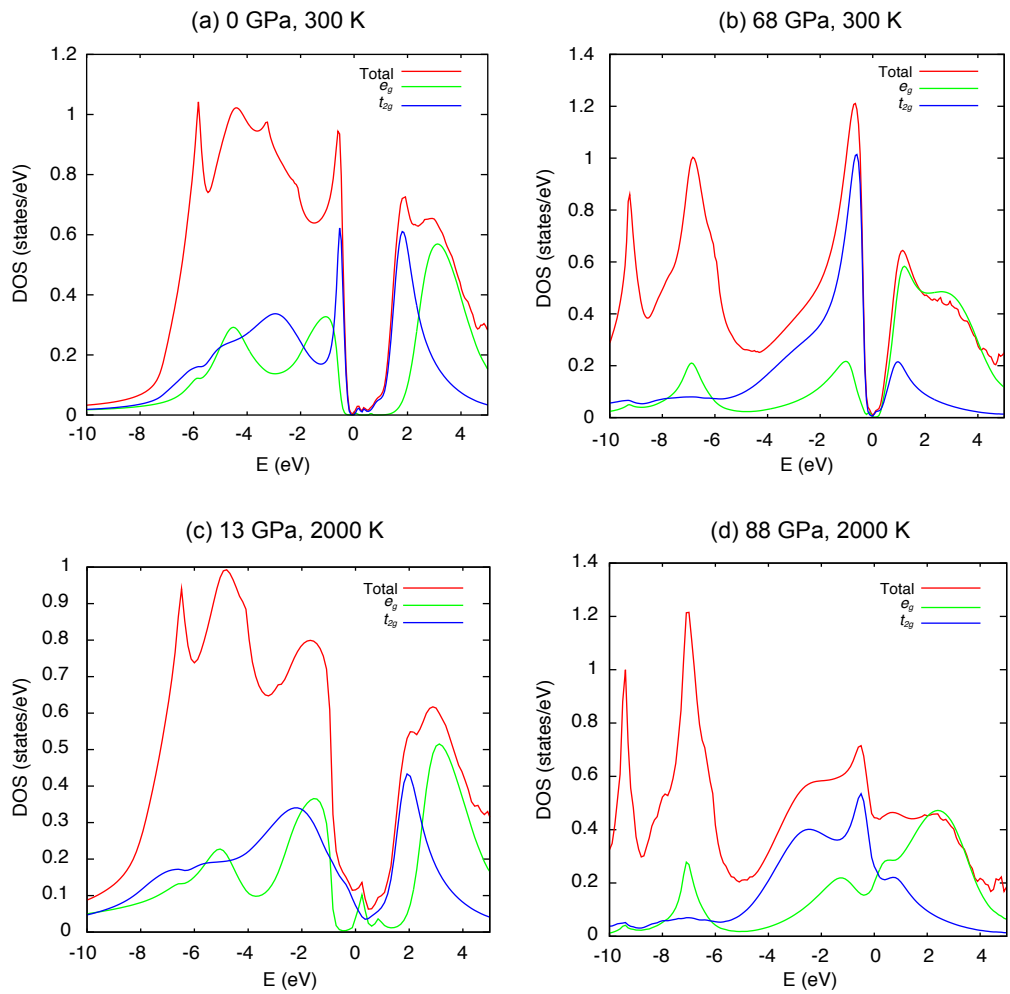

FIG. 3

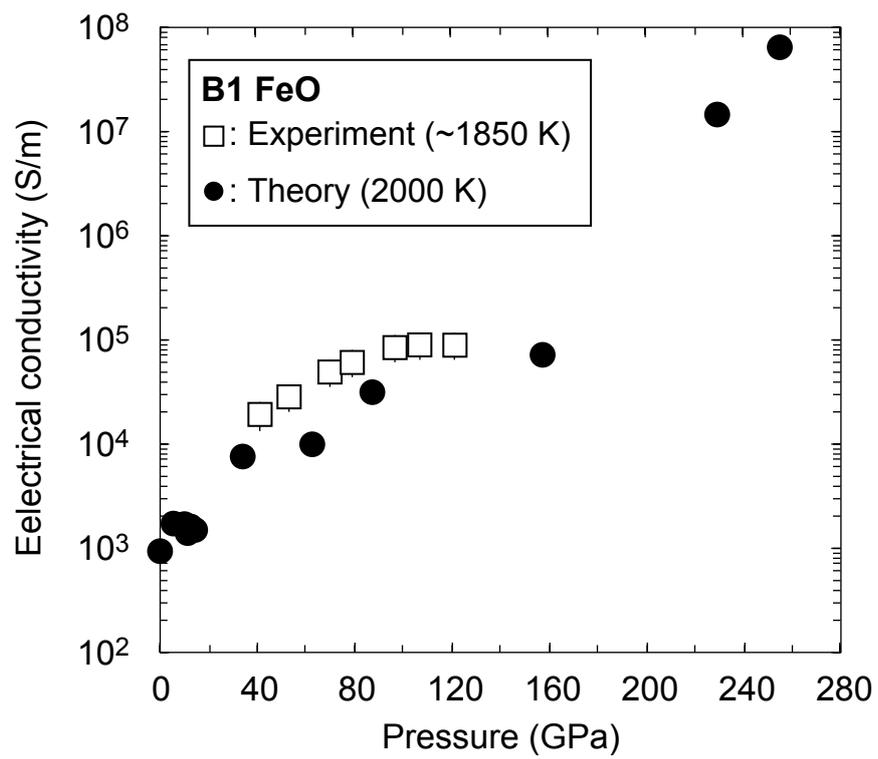

FIG. 4